\documentclass[aps,prb,twocolumn,groupedaddress,showpacs,showkeys]{revtex4}
\usepackage{graphicx}
\begin{document}

\title{Wavefront solution in extended quantum circuits with charge discreteness }
\author{J. C. Flores}
\email{cflores@uta.cl}
\author{Mauro Bologna}
\email{mbologna@uta.cl}
\author{K. J. Chand\'{\i}a}
\email{kchandia@uta.cl}
\affiliation{Departamento de F\'{\i}sica, Universidad de Tarapac\'a, 
Casilla 7-D, Arica, Chile }
\author{Constantino A. Utreras D\'{\i}az }
\email{cutreras@uach.cl} \affiliation{  Instituto de F\'{\i}sica, 
Facultad de Ciencias, Universidad Austral de Chile, 
Campus Isla Teja s/n, Casilla 567, Valdivia, Chile}


\date{March, 2006}

\begin{abstract}
A wavefront solution for quantum (capacitively coupled) transmission lines
with charge discreteness (PRB {\bf 64}, 235309 (2001)) is proposed for
the first time. The nonlinearity of the system becomes deeply related to
charge discreteness. The wavefront velocity is found to depend on a step
discontinuity on the (pseudo) flux variable, $f$, displaying allowed and
forbidden regions (gaps), as a function of $f$. A preliminary study of the
stability of the solutions is presented. The dual transmission line hamiltonian is proposed and
finally, we find a connection with the (quantum) Toda lattice.
\end{abstract}
\maketitle

\section{Introduction}

\label{INTROD}

Today, the broad field of nanostructure is at the heart of many
technological devices~\cite{DATTA,DELOSSANTOS,DELOSSANTOS2,FERRY,HEISS}.
Naturally, at this scale and for low temperature, quantum mechanics plays a
fundamental role~ \cite{FERRY}. Recently, much effort has been devoted to
study nanostructures, using as model that of quantum circuits with
charge discreteness \cite
{LICHEN,CHEN,FLORES,FLORESUTRERAS,UTRERASFLORES,FLORES2,LU}. Work has been 
published on subjects such as persistent currents~ \cite{LICHEN,CHEN}, coupled 
quantum circuits \cite{FLORES,FLORESUTRERAS}, the electronic resonator~\cite{UTRERASFLORES},
quantum point contacts~\cite{FLORES2}, and others \cite{DELOSSANTOS}. In this article, 
we are interested in spatially extended quantum circuits (transmission lines) with 
charge discreteness. This is an interesting specific theoretical subject of broad 
potential applications, since nano-devices could be put together forming chains, and then 
be viewed as electric transmission lines. For instance, electric transport properties in
DNA have been measured recently~\cite{PORATH,HJORT,FINK}; some degree of
disagreement related to conducting properties exists, nevertheless, it is 
clear that the DNA molecule could be viewed and modeled as a quantum transmission line. 
Moreover, molecular electronic circuits~\cite{AVIRAM} are actively studied 
theoretically and experimentally. In such systems, chains of individual molecules 
form a line of circuits. Therefore, quantum circuits is a broad field involving 
future applications from the perspective of nano-devices, electric transmission in 
macromolecules, left-handed materials, Toda lattice, and others.

In this work we will consider a wavefront solution for an extended quantum
circuit (transmission line). For the first time, a sequence of bands and
gaps is found and characterized for this specific system with charge
discreteness, giving hopes for the description of more complex extended
systems (dual transmission lines, complex basis, etc.). 

In section~\ref{SecTrans} we will introduce the Hamiltonian for coupled circuits, and
the equations of motion for the spatially continuous systems. In Sec.
~\ref{WAVEFRONT}, the wavefront solution is obtained and the band-gap structure
is characterized, in Sec.~\ref{STABILITY} the stability question is
considered. In Sec.~\ref{DUAL} the dual line hamiltonian is introduced. In
Sec.~\ref{TODA} the connection with Toda lattice is presented. Finally, we
state our conclusions.

\section{Quantum Transmission lines with discrete charge }
\label{SecTrans}

It is known that for a chain of quantum capacitively-coupled quantum
circuits with charge discreteness ($q_{e}$), the Hamiltonian may be written
as~\cite{FLORES}:

\begin{equation}
\widehat{H}=\sum\limits_{l=-\infty }^{\infty }\left\{ 
\frac{2\hbar ^{2}}{Lq_{e}^{2}}\sin ^{2}(\frac{q_{e}}{2\hbar } \widehat{\phi }_{l})+\frac{1}{2C}
\left( \widehat{Q}_{l}-\widehat{Q}_{l-1}\right) ^{2}\right\} ,
\label{EqHamilton}
\end{equation}
\newline
where the index $l$ describes the cell (circuit) at position $l$, containing
an inductance $L$ and capacitance $C$. The conjugate operators, charge $\widehat{Q}$ 
and pseudoflux $\widehat{\phi },$ satisfy the usual commutation rule 
$[\widehat{Q}_{l},\widehat{\phi }_{l^{\prime }}]=i\hbar \delta _{l,l^{\prime }}$. A 
spatially extended solution of Eq. (\ref{EqHamilton}) will be called a {\em cirquiton like solution},
corresponding to the quantization of the classical electric transmission
line with discrete charge (i.e. elementary charge $q_{e}$). Note that in the
formal limit $q_{e}\rightarrow 0$ the above Hamiltonian gives the well-known
dynamics related to the one-band quantum transmission line, similar to the
phonon case. The system described by Eq. (\ref{EqHamilton}) is very
cumbersome since the equations of motion for the operators are highly
nonlinear due to charge discreteness. However, this system is invariant
under the transformation $Q_{l}\rightarrow \left( Q_{l}+\alpha \right) $,
that is, the total pseudoflux operator 
$\widehat{\Phi }=\sum \widehat{\phi }_{l}$ commutes with the Hamiltonian; in turn, 
the use of this symmetry helps us in simplifying the study of this system.

To handle the above Hamiltonian, we will assume a continuous approximation
(infrared limit); that is to say, we shall use $\left( Q_{l}-Q_{l-1}\right)
\approx {\partial Q}/{\partial x}$~\cite{REMOISSENET}. In this way, it is
possible to re-write the sum of Eq. (\ref{EqHamilton}) as the integral $
\sum\limits_{l=-\infty }^{\infty }\approx \int\limits_{-\infty }^{\infty }dx$,
 where $x$ is a dimensionless variable used to denote the position in the
chain (i.e. the equivalent of  $l$), since we have not introduced so far the
cell size of the circuit. In this approximation, the Hamiltonian becomes

\begin{equation}
\widehat{H}=\int\limits_{-\infty }^{\infty }\widehat{\mathcal{H}}
~dx=\int\limits_{-\infty }^{\infty }\left\{ \frac{2\hbar ^{2}}{Lq_{e}^{2}}
\sin ^{2}(\frac{q_{e}}{2\hbar }\widehat{\phi })+\frac{1}{2C}\left( 
\frac{\partial \widehat{Q}}{\partial x}\right) ^{2}\right\} dx,  
\label{EqContinuo}
\end{equation}
where $\widehat{\mathcal{H}}$ represents the Hamiltonian density operator
for the fields \ $\widehat{\phi }(x)$ and $\widehat{Q}(x)$ and where 
$[\widehat{Q}(x),\widehat{\phi }(x^{\prime })]=i\hbar \delta _{x,x^{\prime }}$. 
From the above Hamiltonian we find the equations of motion (Heisenberg
equations):

\begin{eqnarray}  
\label{EqSistema2}
\frac{\partial }{\partial t}\widehat{\phi } &=&\frac{1}{C}\frac{\partial ^{2}}{\partial x^{2}}\widehat{Q}, \\
\label{cuatro}
\frac{\partial }{\partial t}\widehat{Q} &=&\frac{\hbar }{Lq_{e}}\sin ( \frac{q_{e}}{\hbar }\widehat{\phi }) .
\end{eqnarray}

\section{Wavefront solutions}

\label{WAVEFRONT}

As stated previously, we are interested in wavefront-like solutions of the
system of Eqs.(\ref{EqSistema2}-\ref{cuatro}). We proceed in the standard way \cite{REMOISSENET}, by
assuming travelling wave solutions for our operators, i.e., we define a new
variable $z=x-vt$, and assume

\begin{eqnarray}
\widehat{Q}(x,t) &=&\widehat{Q}(z), \\
\widehat{\phi }(x,t) &=&\widehat{\phi }(z),
\end{eqnarray}
where $v$ stands for the unknown propagation velocity with dimensions of
inverse time, since we did not introduce a grid length parameter in our
continuum approximation. Therefore, from the Heisenberg equations of motion
(\ref{EqSistema2}-\ref{cuatro}), the wavefront equations in the new variable $z$ become

\begin{eqnarray}  
\label{siete}
-v\frac{d}{dz}\widehat{\phi }&=&\frac{1}{C}\frac{d^{2}}{dz^{2}}\widehat{Q},\\
\label{ocho}
-v\frac{d}{dz}\widehat{Q}&=&\frac{\hbar }{Lq_{e}}\sin \frac{q_{e}}{\hbar } \widehat{\phi }.
\end{eqnarray}

From the above pair of equations we obtain a closed equation for the
pseudoflux operator resulting in the "eigenvalue" problem: 
\begin{equation}
\frac{\hbar }{LCq_{e}}\sin \frac{q_{e}}{\hbar }\widehat{\phi }(z)=v^{2}
\widehat{\phi }(z),  \label{eigenvalue}
\end{equation}
where the integration constant has been chosen as zero by simplicity (however, see eq.~\ref{Eq14}). 
Equation (\ref{eigenvalue}) corresponds to an eigenvalue problem for the non-linear
super-operator $\mathcal{L}(\widehat{\phi })=\sin (\widehat{\phi })$. There
is at least two kinds of solutions: a projection operator $\widehat{\phi }=
\widehat{P}$, satisfying $\widehat{P}^{2}=\widehat{P}$, and $\widehat{\phi }=
\widehat{\sigma }$ satisfying $\widehat{\sigma }^{2}=1$. For simplicity, we
shall consider only the first case (a projector). Consider the pseudoflux
operator only in one LC-cell of the chain, and its spectral decomposition in
the Schr\"{o}dinger picture $\widehat{\phi }_{cell}=\int \phi \left| \phi
><\phi \right| d\phi $. Now, pick-up only one term from there (call it $\phi
_{0}$, say) and consider now the well-defined operator

\begin{equation}
\label{Eqdiez}
\widehat{\phi }=f\widehat{P}_{0} ~~{\rm  where }~~ \widehat{P}_{0}=\left| \phi
_{0}><\phi _{0}\right| ,
\end{equation}
where $f$ is an arbitrary pseudoflux parameter, and replace into Eq.(~\ref
{eigenvalue}). At this point, a note concerning the validity for the commutation
rules may be in order: the projector operator $\widehat{P}_{0}$ defines a one dimensional subspace
in which subspace the commutation rule between $\widehat{\phi }=f\widehat{P}_{0}$ 
and the projected charge operator $\widehat{P}_{0}\widehat{Q}\widehat{P}_{0}$, is verified.

Now, since $\widehat{P}_{0}$ is a projector, then the equation for
the pseudoflux $f$ becomes related to the velocity by

\begin{equation}
\label{Eqonce}
v^{2}=\frac{1}{LC}\frac{\sin (q_{e}f/\hbar )}{(q_{e}f/\hbar )}.
\end{equation}
Since both signs ($\pm f$) are possible, then we can construct the wavefront
solution (step $2f$) of the equations of motion (\ref{EqSistema2}-\ref{cuatro}):

\begin{equation}
\widehat{\phi }_{sol}(z)=\left\{ 
\begin{array}{cc}
+f\widehat{P}_{0}, & z>0 \\ 
-f\widehat{P}_{0}, & z<0
\end{array}
\right. ,  
\label{sol}
\end{equation}
corresponding to a solution with zero total flux (Sec. \ref{INTROD}).
Concerning the matching condition at $z=0$, the solution (\ref{sol}) is in
complete agreement with the matching implying explicitly Eqs.(\ref{siete}-\ref{ocho}).

\begin{figure}
\label{Figura1}
\includegraphics[width=6.0 cm,angle=-90]{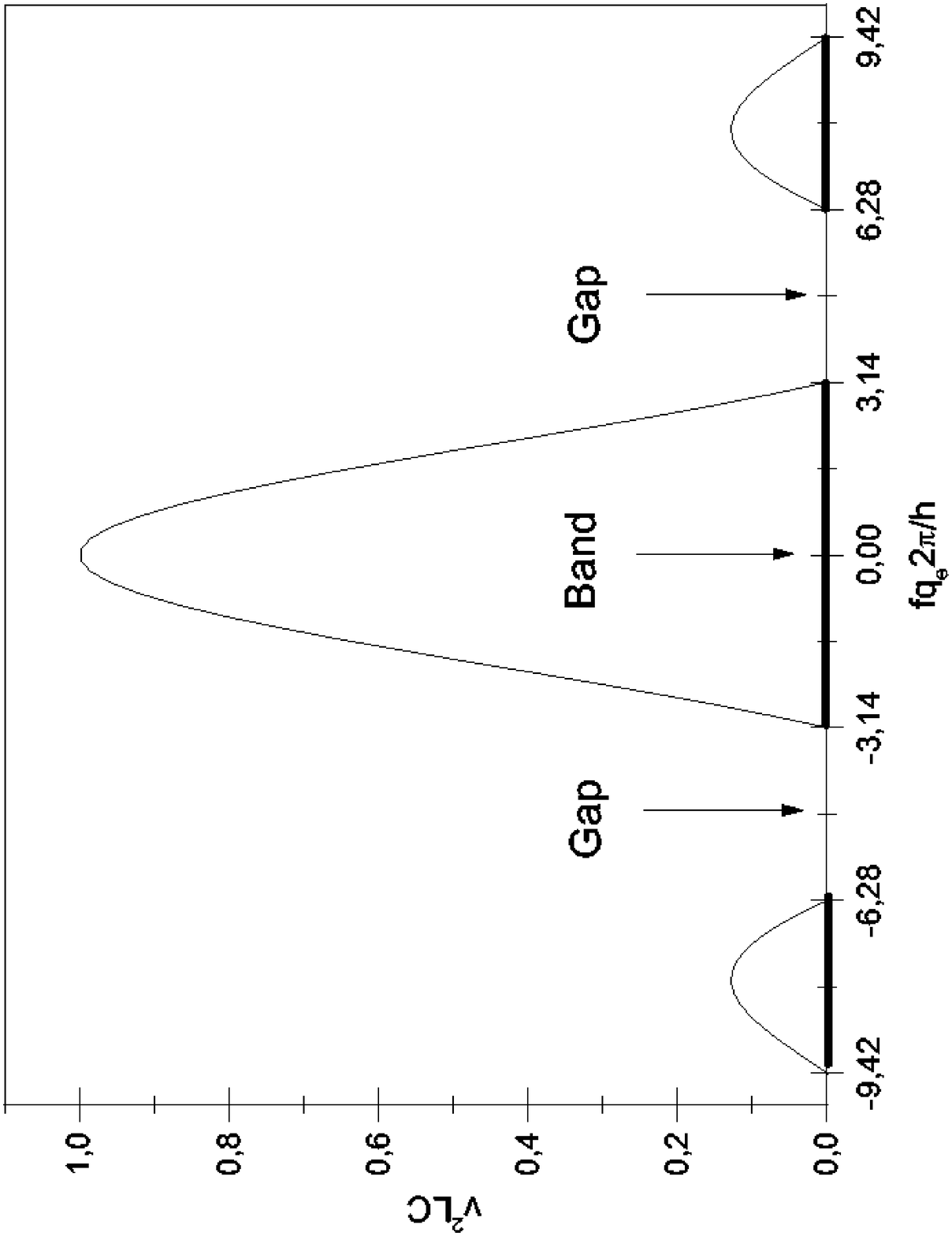}%
\caption{\label{velocidad} Plot of the velocity of the wavefront, as a function of the
flux parameter $f$. As specified by Eq.(~\ref{Eqonce}) there is a structure of bands and gaps. 
The main allowed velocity band is twice as wide as the other allowed bands. This structure 
is a direct consequence of charge discreteness, an effect that dissapears in the limit $q_{e}\rightarrow 0$. }
\end{figure}

The condition $v^{2}\geq 0$ on the wave-front velocity gives the band-gap
conditions on the system. In fact, from (\ref{Eqonce}) the restriction 
\begin{equation}
\frac{\sin \frac{q_{e}}{\hbar }f}{\frac{q_{e}}{\hbar }f}\geq 0,
\label{restriction}
\end{equation}
means that there exists a sequence of bands and gaps. The figure~\ref{velocidad}
shows a plot of the wavefront velocity (\ref{Eqonce}) for different values of the
pseudoflux $f$, in which it is seen an alternating sequence of bands and gaps,
corresponding to propagating and forbidden modes. Note that the main
allowed band ($-\pi \hbar /q_{e}<f<\pi \hbar /q_{e}$) is twice as wide as
the other allowed bands. 

Recall that the integration constant was set equal to zero to obtain 
(\ref{eigenvalue}), now consider the nonzero case, then the wavefront velocity
becomes formally 
\begin{equation}
\label{Eq14}
v^{2}\left( \widehat{\phi }-\widehat{C}\right) =\frac{\hbar }{LCq_{e}}
\left(\sin \frac{q_{e}}{\hbar }\widehat{\phi }-\sin \frac{q_{e}}{\hbar }\widehat{C}
\right) .
\end{equation}
where $\widehat{C}$ is an integration constant.

\section{Stability}

\label{STABILITY}

We present some preliminary results concerning the stability of the solution~(\ref{Eqdiez}); a 
more detailed treatment is currently under consideration. Consider now the perturbative
solutions 
\begin{eqnarray}
\widehat{\phi } &=&\widehat{\phi }_{sol}+\widehat{\varepsilon }  \label{diez}
\\
\widehat{Q} &=&\widehat{Q}_{sol}+\widehat{\eta },  \nonumber
\end{eqnarray}
where the operators $\widehat{\varepsilon }$ and $\widehat{\eta }$ are the
perturbation and then, assumed with small eigenvalues. Moreover, consider
the well known perturbative expansion (\cite{KUBO})

\begin{equation}
\sin \left( \widehat{\phi }+\widehat{\varepsilon }\right) =\sin \left( 
\widehat{\phi }\right) +{Re}\left( e^{i\widehat{\phi }}\int_{0}^{1}d\theta
~e^{-i\widehat{\phi }\theta }~\widehat{\varepsilon }~e^{i\widehat{\phi }
\theta }\right) .
\end{equation}
Since $\widehat{\phi }=f\widehat{P}_{0},$ and as said $\widehat{P}_{0}$ is a
projector, then from the above equation (and (3-4)) we have the linear
evolution equation for the perturbation $\widehat{\varepsilon }$:

\begin{equation}
LC\frac{\partial ^{2}}{\partial t^{2}}\widehat{\varepsilon }= Re \frac{
\partial ^{2}}{\partial x^{2}}\left( e^{i\frac{qf}{\hbar }\widehat{P}
_{o}}\int_{0}^{1} ~d\theta~e^{-i\frac{qf}{\hbar }\widehat{P}_{o}\theta } ~ \widehat{
\varepsilon }~ e^{+i\frac{qf}{\hbar }\widehat{P}_{o}\theta }\right) .
\end{equation}

We shall consider two cases:

\begin{enumerate}
\item[(a)]  $\widehat{\varepsilon }=\varepsilon (x,t)\widehat{P}_{0}$. Here 
$ [\widehat{\varepsilon },\widehat{P}_{0}]=0$ and then we have the linear
wave equation $LC\frac{\partial ^{2}}{\partial t^{2}}\varepsilon =\left(
\cos \frac{qf}{\hbar }\right) \frac{\partial ^{2}}{\partial x^{2}}
\varepsilon $ and the perturbations are unstable when 
$\left( \cos \frac{qf}{\hbar }\right) <0$.

\item[(b)] $\widehat{\varepsilon }=\varepsilon (x,t)(\widehat{I}-\widehat{
P}_{0}).$ In this case the wave equation becomes $LC\frac{\partial ^{2}}{\partial t^{2}}\varepsilon = 
\frac{\partial ^{2}}{\partial x^{2}}\varepsilon $, and the perturbation is stable.
\end{enumerate}
\vskip 0.5 cm

\section{ Dual transmission line}

\label{DUAL}

It is well known that the direct classical L-C transmission line (related to
equation~\ref{EqHamilton}) has associated a dual transmission line: in the direct 
line the interaction between cells is through capacitances, while in the dual it is through
inductances. The dual line is closely related to the so-called {\em left-handed}
materials, and then its quantization is actually important. Moreover, the
role of charge discreteness must be also considered. This two step process
(quantization and charge discreetness) could be realized in analogy with the
direct line (section~\ref{INTROD}) but in this case long range interactions
between cells appear in the Hamiltonian. In fact, the expression for the
Hamiltonian is (\cite{CHANDIA}):

\begin{widetext}
\begin{equation}
\widehat{H}=\frac{\hbar ^{2}}{2\pi Lq_{e}^{2}}\sum\limits_{l,n}\left( \int
dk \frac{e^{ik(l-n)}}{1-\cos k}\sin \frac{q_{e}}{2\hbar }\widehat{\phi }
_{l}\sin \frac{q_{e}}{2\hbar }\widehat{\phi }_{l}\right) + \sum_{l}\frac{
\widehat{Q}_{l}^{2}}{2C}.
\end{equation}
\end{widetext}

The equation of motion the charge and pseudoflux may be obtained after some
algebra. We shall not continue the description of the dual line, but we
shall only mention here that the expressions are quite involved and, so far,
no explicit solution have been found.

\section{Toda Lattice and cirquitons}

\label{TODA}

The cirquiton Hamiltonian (\ref{EqHamilton}) gives the equations of motion
for charge \ and pseudo-flux in the cell $l:$

\begin{equation}
\frac{\partial }{\partial t}\widehat{\phi }_{l}=\frac{1}{C}\left( \widehat{Q}
_{l+1}+\widehat{Q}_{l-1}-2\widehat{Q}_{l}\right) ,
\end{equation}

\begin{equation}
\frac{\partial }{\partial t}\widehat{Q}_{l}=\frac{\hbar }{Lq_{e}}\sin \frac{
q_{e}}{\hbar }\widehat{\phi }_{l}.
\end{equation}
Equations (\ref{EqSistema2}-\ref{cuatro}) in section \ref{INTROD} are the spatially continuous version
of the above pair of equations. Then, for the pseudo-flux operator we have
the closed non linear equation of motion: 
\begin{equation}
LC\frac{\partial ^{2}}{\partial t^{2}}\widehat{\phi }_{l}=\frac{\hbar }{q_{e}
}\left( \sin \frac{q_{e}}{\hbar }\widehat{\phi }_{l+1}+\sin \frac{q_{e}}{
\hbar }\widehat{\phi }_{l-1}-2\sin \frac{q_{e}}{\hbar }\widehat{\phi }
_{l}\right) ,  \label{cirquiton}
\end{equation}
which, in the formal limit $q_{e}\rightarrow 0$ gives the usual one-band
system with frequency spectrum $\omega (k)=\frac{2}{\sqrt{LC}}\left| \sin
(k/2)\right| $, as expected. Consider now the equation:

\begin{equation}
LC\frac{\partial ^{2}}{\partial t^{2}}\widehat{\phi }_{l}=\frac{\hbar }{
iq_{e}}\left( e^{\frac{iq_{e}}{\hbar }\widehat{\phi }_{l+1}}+e^{\frac{iq_{e}
}{\hbar }\widehat{\phi }_{l-1}}-2e^{\frac{iq_{e}}{\hbar }\widehat{\phi }
_{l}}\right) ,  \label{qtoda}
\end{equation}
then any hermitian solution of this last equation is also solution of the
cirquiton equation (\ref{cirquiton}). To show this, it suffices to conjugate
Eq. (\ref{qtoda}) and subtracting. Notice that any hermitian solution of Eq.(\ref{qtoda}) 
is also a solution of (\ref{cirquiton}), but the inverse is not necessarily true.

On the other hand, the broad field related to the Toda lattice is studied in
depth in a variety of branches in physics \cite{REMOISSENET} including
non-linear physics, statistical mechanics, classical electric circuits, etc.
 The classical evolution equation for the Toda lattice has the generic form: 
 
\begin{equation}
M\frac{\partial ^{2}}{\partial t^{2}}\phi _{l}=-A\left( e^{-B\phi
_{l+1}}+e^{-B\phi _{l-1}}-2e^{-B\phi _{l}}\right)  \label{ctoda}
\end{equation}
where $M,A$ and $B$ are real constant. So, we have the important result: the
formal replacement $q_{e}\rightarrow iq_{e}$ transforms (\ref{qtoda}) into
the (quantum ) Toda lattice equation. This makes a direct connection between
Toda lattice and cirquiton theory.

\section{ Final Remarks}

For the quantum electric transmission line with charge discreteness
described by the Hamiltonian Eq.(\ref{EqHamilton}), and equations of motion 
(\ref{EqSistema2}-\ref{cuatro}), a one parameter ($f$) wavefront solution was 
found (Eqs. \ref{Eqonce}-\ref{sol}). The condition Eq.(\ref{Eqonce})
on the velocity generates a band-gap structure dependent on the
pseudoflux parameter $f$ (see figure~\ref{velocidad}), namely, there exist regions 
(values of $f$) for which a solitary wavefront propagates with constant speed. 
The existence of the band-gap structure described is the main result of this work.

\begin{acknowledgments}

J.C. FLores, K. J.Chandia and C.A. Utreras-D\'{i}az acknowledge financial
support by FONDECYT Grant \# 1040311. C.A. Utreras-D\'{i}az also
acknowledges support from Universidad Austral de Chile (DID Grant \#
S-2004-43). Useful discussions originally were carried-out with V. Bellani
(Pavia University) and P. Orellana (UCN).
\end{acknowledgments}

\end{document}